\documentclass[%
 reprint,
superscriptaddress,
 amsmath,amssymb,
 aps
]{revtex4-2}

\usepackage{graphicx}
\usepackage{dcolumn}
\usepackage{bm}
\DeclareMathAlphabet{\mathpzc}{OT1}{pzc}{m}{it}
\usepackage[colorlinks=true, allcolors=blue]{hyperref}

\usepackage[caption=false]{subfig}
\usepackage{amssymb}
\usepackage{amsmath}

\usepackage{graphicx,bm}
\usepackage{verbatim}

\usepackage{float}

\begin{document}


\title{Effect of temporary lockdowns on disease extinction risk in assortative networks}

\author{Elad Korngut} \author{Michael Assaf}

\affiliation{Racah Institute of Physics, Hebrew University of Jerusalem, Jerusalem 91904, Israel }

\begin{abstract}
Changing environmental conditions can significantly affect the dynamics of disease spread. These changes may arise naturally or result from human interventions; in the latter case, lockdown measures that lead to abrupt but temporary reductions in transmission rates are used to combat disease spread.  Yet, the impact of these measures on rare events in \textcolor{black}{heterogeneous} populations \textcolor{black}{remains understudied}. Here, we analyze the susceptible-infected-susceptible (SIS) model in a stochastic setting where disease extinction---a sudden clearance of the infection—--occurs via a rare, large fluctuation. We use a semiclassical approximation and numerical simulations on  heterogeneous assortative networks, with degree-degree correlations between neighboring nodes, to show how the extinction risk of the disease depends on the lockdown's duration and magnitude, \textcolor{black}{and on the network topology.}
\end{abstract}

\maketitle

\section{\label{sec:Intro}Introduction}
The spread of infectious diseases through populations is largely determined by contact patterns between individuals and the rate of transmission across those contacts \cite{Hethcote_2000_Mathematics_of_Infectious_Diseases,Pastor-Satorras_2001_Epidemic_Spreading_in_Scale_Free_Networks,Pastor-Satorras_2001_Epidemic_dynamics_and_endemic_states_in_complex_networks,Pastor-Satorras_2002_finite_size_scale-free_networks,keeling2005networks,Dorogovtsev_2008_Critical_complex_networks,keeling_2011_modeling,Pastor-Satorras_2015_anneald_net}. During the COVID-19 pandemic, widespread quarantine measures, such as lockdowns and mandatory mask-wearing, were implemented to curb transmission \cite{Block2020,Hellewell2020}. These interventions led to abrupt and substantial changes in both contact patterns and transmission rates. While many studies explored the impact of periodically varying or fluctuating environmental conditions, more recently, in response to various large-scale pandemics, attention has shifted toward understanding the effects of temporary lockdown measures \cite{Keeling_2008_stochastic_dynamics,shaw2008_adaptive_networks,Assaf2008time-modulated,assaf2009population,bacaer2015stochastic,assaf2013cooperation,Israeli2020switching,Donohue2020,Block2020,Hellewell2020}.

To describe  disease spread reliably, epidemiological models must incorporate both the underlying contact structure and the dynamics of disease progression. The latter is typically captured using compartmental models, which divide the population into distinct compartments, each representing a possible state of an individual. These states commonly include susceptible individuals, infected individuals who carry the disease, and recovered individuals who have acquired immunity. In the absence of long-term immunity, however,  the susceptible-infected-susceptible (SIS) model, which includes only two compartments of susceptibles (S) and infected (I), provides an adequate description of the dynamics~\cite{Pastor-Satorras_2001_Epidemic_Spreading_in_Scale_Free_Networks,Pastor-Satorras_2001_Epidemic_dynamics_and_endemic_states_in_complex_networks,keeling2005networks,Dorogovtsev_2008_Critical_complex_networks,keeling_2011_modeling}. Here, susceptible individuals transition to the infected compartment upon infection, while infected individuals return to the susceptible compartment upon recovery.

Even in its simplest formulation, the SIS model exhibits nontrivial long-term behavior.  When the infection rate exceeds a critical threshold (discussed below), the system approaches a persistent endemic state in the infinite-population limit, with the disease surviving indefinitely \cite{Pastor-Satorras_2015_anneald_net}. However, for large (yet finite) populations, this state becomes a long-lived metastable state, with random demographic fluctuations ultimately leading to disease extinction~\cite{dykman_1994_Large_fluctuations}.  In this case, the quantities of interest are the  extinction time and extinction probability (EP) of the disease. Computation of the EP---the probability of escaping
the long-lived endemic state via a large fluctuation---is simplified under the assumption that for a large population size, the metastable endemic state is quasistationary, and slowly decays in time~\cite{dykman_1994_Large_fluctuations,assaf_2010_Extinction,assaf2011fixation}. Here, both the quasistationary distribution (QSD) around the endemic state and the EP can be  analytically computed, in well-mixed environments, in which each individual  interacts uniformly with all others~\cite{Hethcote_2000_Mathematics_of_Infectious_Diseases,keeling_2011_modeling,NOLD1980,ovaskainen_2001,assaf_2010_Extinction,assaf2011rare,assaf_2017_WKB}. 

However, predicting rare events in more realistic environments---where contact patterns shape disease transmission---remains a  challenge, due to the high dimensionality and complex coupling between degrees of freedom~\cite{hindes_2016_paths}. In such cases, networks are used to describe the underlying topology of contacts, where nodes represent individuals and links capture interactions between them. 
In static networks with a fixed structure, degree heterogeneity plays a key role in disease dynamics and has a profound effect on rare events statistics~\cite{clancy_2013_spread_of_infection,clancy_2018_heterogeneous_populations,clancy_2018_persistence,hindes_2019_undirected_degree_hetro,korngut_2022_direction,korngut_2025_we}. Recently, assortative networks---where nodes with similar degrees tend to connect \textcolor{black}{to each other}~\cite{newman_2002_asso_prl,Pastor-Satorras_2015_anneald_net,Silva2019_assortative_table}---have been shown to alter the extinction risk, even when the degree distribution is  fixed~\cite{korngut2025_assortative}. 

In this work we are interested in exploring the effect of \textcolor{black}{varying the disease parameters in time}, as means of mitigating and controlling the epidemic. Such scenarios have been considered in well-mixed settings by taking time-varying infection and recovery rates~\cite{Keeling_2008_stochastic_dynamics,Assaf2008time-modulated,assaf2009population,shaw2008_adaptive_networks,hindes2023outbreak,bacaer2015stochastic}. In more complex topologies of heterogeneous  networks, reducing exposure and mitigating outbreaks may be achieved by individual link severing or rewiring, or more global measures such as decreasing the overall number of links due to  lockdowns or seclusions~\cite{gross2006epidemic,shaw2008_adaptive_networks,Shaw2010enhanced,Crokidakis_2012,Tunc2013,Juher2013,RIZZO2016_Ebola,DiDomenico2021,DiDomenico2021Curfew,Du_2021_cost,Meidan2021,Atias2025_quarantine}.

An important example of \textcolor{black}{varying the  parameters in time, which has a dramatic effect on the epidemic's outcome,} arises from the implementation of \textit{temporary} lockdowns, causing an abrupt reduction in the infection rate for a prescribed period. Such lockdowns have been studied in terms of the trade-off between their cost---in duration and stringency---and their effectiveness in mitigating disease spread~\cite{Meidan2021,Atias2025_quarantine}. 
Moreover, it has been shown that the way the lockdown is carried out strongly influences the epidemic dynamics; for example, intermittent lockdowns, where distinct fractions of the population are quarantined periodically, have been shown to enhance containment and lead to disease eradication~\cite{Meidan2021,Hindes2021_risk,SEJUNTI2024}. 
Yet, the question of how such quarantines or lockdowns can be optimized  in the presence of contact heterogeneity and demographic fluctuations, which drive the disease towards extinction, \textcolor{black}{has received little attention so far}.

Here, we study the effect of temporary lockdown measures on the EP of the disease. The lockdown is modeled as a brief, sharp reduction in the infection rate occurring in an otherwise endemic population, due to intervention measures such as increased seclusion, followed by a return to the baseline transmission level. We compute the EP in a homogeneous setting using a semi-classical approximation, and perform extensive numerical simulations to unravel the interplay and tradeoff between the lockdown duration and magnitude,  and network's structure  manifested by its extent of heterogeneity and assortativity.

\section{\label{sec:theory} Theoretical Formulation}
We begin by formulating the theoretical SIS model on a heterogeneous  network. In the following we will focus on assortative networks in which  high-degree nodes tend to connect to other high-degree nodes and vice versa~\cite{newman_2002_asso_prl}. The complementary scenario of disassortative networks, in which high-degree nodes tend to connect to low-degree nodes, will be discussed in Sec.~III. In this model, an isolated population of $N$ individuals is divided into two compartments: susceptible ($S$) and infected ($I$). Transitions between the two compartments are possible  via infection, $S+I\to I+I$ at a rate $\beta$ per encounter, or recovery, $I\to S$, at a rate $\gamma$ per individual~\cite{Bailey_1975_book,Pastor-Satorras_2015_anneald_net}. 

In a population network, individuals are represented as nodes and their interactions as undirected links, encoded in the adjacency matrix \( \mathbf{A} \).
Under the annealed network approximation---a mean-field average over an ensemble of networks---the adjacency matrix \( \mathbf{A} \) may be replaced by its expectation value \( \langle \mathbf{A} \rangle \), \textcolor{black}{which satisfies $\langle A_{ij} \rangle=k_jP(k_i|k_j)/[N P(k_i)]$}~\cite{Pastor-Satorras_2015_anneald_net}. Here, interactions between individuals depend on \( P(k'|k) \)---the conditional probability  that a node of degree \( k \) is connected to a node of degree \( k' \), and on the degree distribution \( P(k) \), which specifies the fraction of nodes with degree \( k \), such that the number of degree-\( k \) nodes is \( N_k = N P(k) \), with \( \sum_k N_k = N \). For concreteness, we focus on a specific form of $P(k'|k)$ to account for assortativity~\cite{newman_2002_asso_prl,moreno2003epidemic,Pastor-Satorras_2015_anneald_net,leibenzon2024heterogeneity}
\begin{equation}
\label{eq:conditional_prob}
    P(k'|k) = \frac{k'P(k')}{\left<k\right>}(1-\alpha)+\alpha \delta_{k,k'}.\end{equation}
Here $0\leq\alpha \leq 1$ represents the correlation  between degrees of neighboring nodes, such that a fraction \( \alpha \) of the links tend to connect nodes of the same degree. Note that, this framework is not suitable for disassortative networks ($\alpha<0$), which will be discussed in Sec.~III~\cite{leibenzon2024heterogeneity,korngut2025_assortative}.

Using the conditional probability~(\ref{eq:conditional_prob}), one can write the rate of infection of susceptibles residing on a degree-$k$ node. This rate equals the probability that a degree-$k$ node is connected to a degree-$k'$ node multiplied by the node's degree $k$ and by the fraction of infected on degree-$k'$ nodes. Conversely, the recovery rate of an infected residing on a degree-$k$ node equals the fraction of degree-$k$ infecteds multiplied by $\gamma$. This yields
\begin{eqnarray*}
\label{eq:transition_rate_k_case}
    &\mathbf{I}\xrightarrow{W_{k}^{+}(\mathbf{I})}\textcolor{black} {\mathbf{I}}+\mathbf{1}_k , \quad
    \mathbf{I}\xrightarrow{W_{k}^{-}(\mathbf{I})} \textcolor{black}{\mathbf{I}}-\mathbf{1}_k,
\end{eqnarray*}
where $W_{k}^{+}(\mathbf{I})=\beta \Sigma_{_{k'}}k P\left(k'|k\right)x_{k'}$ is the  infection rate, and $ W_{k}^{-}(\mathbf{I})=\gamma I_k$ is the recovery rate. We have also denoted by $I_k$ the number of infecteds on a degree-$k$ node, $x_k=I_k/N_k$ is their fraction, and  $\mathbf{I}\pm\mathbf{1}_{\!k}$ denotes an increase or decrease by 1 of $I_{k}$.

In the limit of an infinite population size, demographic noise can be neglected, and after a short transient the system converges to a stable endemic state, given by a vector of fractions of infected, $\mathbf{x}=\mathbf{x}^*$. Apart from the endemic stable state, there exists an unstable fixed point at $\mathbf{x}=0$, representing disease extinction or clearance, where the entire population becomes susceptible again.

The existence of the stable endemic state is guaranteed as long as the infection rate $\beta$ is above some critical value $\beta_c$, where a transcritical bifurcation occurs at $\beta=\beta_c$. In the simple case of a well-mixed population of size $N$, one has $\beta_c=\gamma/N$, such that upon defining the basic reproduction number, $R_{0} \equiv \beta / \beta_c=N\beta/\gamma$, bifurcation occurs at $R_0=1$. On the other hand, in the case of a homogeneous network with $k_0$ neighbors per node, one has $R_{0}=k_0\beta/\gamma$, since the critical infection rate is determined only by the close vicinity of each node.

In more complex heterogeneous settings, in order to compute $\beta_c$, one can use the so-called heterogeneous mean-field theory~\cite{Pastor-Satorras_2002_Epidemic_spreading_correlated_networks,Pastor-Satorras_2015_anneald_net}. Here, one computes the connectivity matrix $C_{kk'} = kP(k'|k)$, and the epidemic threshold becomes $\beta_c=\textcolor{black}{\gamma}/\Upsilon^{(1)}$, where $\Upsilon^{(1)}$ denotes the largest eigenvalue of $C_{kk'}$~\cite{Pastor-Satorras_2002_Epidemic_spreading_correlated_networks}. As a result, $R_0$ becomes
\begin{equation}
    R_0 = (\beta/\gamma) \Upsilon^{(1)}.
    \label{eq:reproductive_number_hmf}
\end{equation}
A simple example of a heterogeneous network is given by the  bimodal network, defined by the degree distribution \( P(k) = \left(\delta_{k,k_1} + \delta_{k,k_2}\right)/2 \). This network has two node types: a high-degree node with degree  \( k_1 = k_0(1+\epsilon) \) and a low-degree node with degree \( k_2 = k_0(1-\epsilon) \), where \textcolor{black}{$\epsilon=\sigma/\left<k\right>$} denotes the coefficient of variation (COV) of $P(k)$, and \( \langle k \rangle \) and \textcolor{black}{$\sigma$} are its \textcolor{black}{mean} and \textcolor{black}{standard deviation}. In this case, assuming assortativity strength $\alpha$ and using Eq.~(\ref{eq:conditional_prob}), the connectivity matrix satisfies~\cite{korngut2025_assortative}
\[
\mathbf{C}=\frac{1}{2k_0}\begin{bmatrix}

k_{1}^{2}(1-\alpha)+\alpha k_{1} & k_{1}k_{2}(1-\alpha) \\
k_{1}k_{2}(1-\alpha) & k_{2}^{2}(1-\alpha)+\alpha k_{2}
\end{bmatrix}.
\]
Computing $\Upsilon^{(1)}$ and plugging it into Eq.~\eqref{eq:reproductive_number_hmf}  yields~\cite{korngut2025_assortative}
\begin{eqnarray}
R_0\!=\!\frac{\beta k_0}{2\gamma}\!\left[2 \!-\! (1\!-\!\alpha) (1 \!-\!\epsilon^2) \!+\! \sqrt{ 4 \epsilon^2 \!+\! (1\!-\!\alpha)^2 (1 \!-\! \epsilon^2)^2}\right].
\label{eq:reproductive_bimodal}
\end{eqnarray}
Notably, for $\alpha\to 0$, $R_0=(\beta k_0/\gamma)(1+\epsilon^2)$, which agrees with the general formula, \( R_0 = \beta \langle k^2 \rangle / (\langle k \rangle \gamma) \), valid for random networks with zero assortativity~\cite{Pastor-Satorras_2015_anneald_net}.

For more complex assortative networks, such as having a gamma distribution for the degrees, obtaining an analytical expression for \( \Upsilon^{(1)} \) is generally intractable. Nevertheless, \( \Upsilon^{(1)} \) can be determined numerically. Regardless of the method used to obtain \( \Upsilon^{(1)} \), to maintain a constant \( R_0 \) across different degree distribution types, one must adjust the ratio \( \beta / \gamma \) such that \( R_0 \) satisfies Eq.~\eqref{eq:reproductive_number_hmf}.

Importantly, accounting for demographic noise (due to the discreteness of individuals), the endemic state becomes metastable~\cite{dykman_1994_Large_fluctuations}, and the system ultimately arrives at a disease-free state with probability one. Here, for a large population (or network) size, $N\gg 1$, the lifetime of the metastable state is very long, and extinction only occurs after the system experiences a sufficiently large fluctuation~\cite{assaf_2010_Extinction,assaf_2017_WKB}. In order to measure the extinction risk of the disease, we define the time-dependent extinction probability (EP), \( \mathcal{P}(t) \), as the probability that the system has reached extinction by time \( t \). The EP  increases with time and asymptotically approaches one as \( t \to \infty \). Notably, \( \mathcal{P}(t) \)  can be related to the mean time to extinction (MTE) $\tau$ in simple time-independent scenarios \textcolor{black}{(see Appendix A)}. While analytical expressions for \( \mathcal{P}(t) \) in the large-population limit  exist for well-mixed populations, computing the EP becomes highly complicated  when degree heterogeneity and assortativity are taken into account. In these cases, analytical results are available only under simplifying assumptions, such as near the bifurcation point or for weakly heterogeneous networks with either directed or undirected links \cite{clancy_2018_heterogeneous_populations,hindes_2019_undirected_degree_hetro,korngut_2022_direction,korngut2025_assortative}.
Naturally, when the rates are explicitly time dependent, this quantity can only be calculated numerically, \textcolor{black}{see Sec.~\ref{sec:numeric}}. 

Having laid down the general theoretical framework, we now turn to the scenario of a temporary lockdown, in which after the system has settled in the long-lived metastable state, a temporary
environmental change occurs. Here, at  $t=t_0$ the environment switches to a new (constant) state with a lower infection rate for a finite
period of time $T$, whereas at time $t_0+T$ the system returns to the original state, with the original infection rate $\beta_0$.  Thus, the time-dependent infection rate reads
\begin{equation}
\label{eq:beta_quant}
\beta(t) = 
\begin{cases}
\beta_0, & \text{for } t < t_0 \text{ or } t > t_0 +  T, \\
\beta_0 (1 - \xi), & \text{for } t_0 \leq t \leq t_0 +  T, 
\end{cases}
\end{equation}
where $0\leq \xi\leq 1$ measures the strength of the lockdown, and $T$ measures its duration. Notably, the new environmental
state is advantageous for disease eradication, as during the lockdown, the EP greatly increases.

The question we address here is how the EP increases due
to this environmental change in the form of a temporary lockdown occurring at $t = t_0$. In fact, for $N\gg 1$, at
times $t < t_0$ (assuming that $t_0$ is much shorter than the MTE $\tau$)
the pre-lockdown EP is expected to be exponentially small, and is given by $\mathcal{P}_b\simeq 1-e^{-t/\tau}$~\cite{Assaf2008time-modulated,assaf2009population,Israeli2020switching}. At times $t_0 < t < t_0 + T$ , when the lockdown is applied, the EP grows at a faster
rate, reaching a post-lockdown value of $\mathcal{P}_a$ at $t = t_0 + T$, 
which satisfies $\mathcal{P}_a\gg \mathcal{P}_b$. We are interested in computing the increase in the EP due to the
lockdown $\Delta \mathcal{P} \equiv \mathcal{P}_a-\mathcal{P}_b$ which approximately equals $\mathcal{P}_a$~\footnote{In sharp contrast to the EP, which significantly grows, the MTE is almost not affected by the lockdown as long as the lockdown duration is sufficiently small~\cite{assaf2009population,Israeli2020switching}.}. In a homogeneous setting $\mathcal{P}_a$ can be found analytically, see Appendix A. In the following we compare the results of  homogeneous and heterogeneous networks to explore the lockdown impact under varying network heterogeneity and assortativity.

It is important to note that, as either the strength or duration of the intervention increases, the resulting impact on the EP becomes more substantial. However, stronger or longer interventions are also more resource-intensive and harder to implement. Since increasing heterogeneity or assortativity leads to a shorter  disease lifetime~\cite{clancy_2018_persistence,hindes_2019_undirected_degree_hetro,korngut_2022_direction,korngut2025_assortative}, we expect that the lockdown effectiveness will increase in \textcolor{black}{heterogeneous} networks compared to a well-mixed setting. 
Our goal is therefore to determine the level of resources a lockdown must allocate to sustainably increase the EP in a realistic heterogeneous setting. In other words, given a desired outcome of the lockdown---i.e., increasing the clearance likelihood by a certain prescribed factor---we wish to  explore how the lockdown duration and strength should vary, as a function of the  network's heterogeneity and assortativity.

\section{\label{sec:numeric}  NUMERICAL METHODOLOGY}

To study the EP in heterogeneous, assortatively mixed populations, we constructed synthetic networks to represent the desired topology. \textcolor{black}{The latter was determined by a prescribed degree distribution \( P(k) \) and prescribed assortativity coefficient $\alpha$.  We then used a kinetic Monte Carlo (KMC) method to simulate the dynamics, while keeping the topology fixed throughout the simulation}.

We generated a structure of an uncorrelated (random) network with zero assortativity using the configuration model~\cite{Fosdick_2018_config_model}. Here, each node is assigned a predefined number of stubs based on its degree, and the stubs are randomly paired to form links. This procedure preserves the degree distribution while eliminating correlations between connected nodes.

To generate assortative and disassortative networks we used the Xulvi-Brunet–Sokolov algorithm~\cite{Xulvi-Brunet_2004_Reshuffling_assortative}. Initially, the network topology is generated using the configuration model. Then, links are iteratively rewired to adjust the assortativity. In each step, two links involving four distinct nodes are randomly selected, and the nodes are sorted by degree. With probability \( |\alpha|\leq 1 \), the rewiring process connects either two high-degree nodes to promote assortativity or a high-degree node to a low-degree node to induce disassortativity, whereas, with probability \( 1 - |\alpha| \), the rewiring is done randomly. If the rewiring step creates links that already exist in the network, the step is repeated to avoid link duplication. This process allows for the generation of both assortative and disassortative networks with arbitrary $-1\leq\alpha\leq1$. \textcolor{black}{To verify consistency, the degree assortativity coefficient was computed for the generated networks using the definition of~\cite{newman_2002_asso_prl}.}

Regardless of the method used to construct the network, the resulting adjacency matrix can be used to compute its largest eigenvalue, which yields the basic reproduction number \( R_0 \) from Eq.~(\ref{eq:reproductive_number_hmf}). The reason we determine $R_0$ is that, throughout our calculations we keep $R_0$ constant\textcolor{black}{, for each network realization,} regardless of the network topology~\footnote{\textcolor{black}{In general, the eigenvalues of the connectivity matrix $\mathbf{C}$ (used under the annealed network approximation) and of the adjacency matrix $\mathbf{A}$, differ. Yet, we checked that for each network realization we have generated (bimodal or gamma), the difference between the largest eigenvalue of $\mathbf{C}$ and $\mathbf{A}$ was $<\!1\%$ due to the absence of major hubs~\cite{Silva2019_assortative_table}}.}. We do so in order to keep the distance to bifurcation constant and eliminate the deterministic effects while studying the dependence of the EP on the network and lockdown parameters.

Once $R_0$ is computed, each node’s infection and recovery rates are computed based on its neighbors. Here, each node \( j \) can be in one of two states, \( j_S \) (susceptible) or \( j_I \) (infected), with transitions between the two  simulated using the Gillespie algorithm~\cite{GILLESPIE1976_second_paper,Gillespie_1977}. The transitions, $j_S\to j_I$ and $j_I\to j_S$ occur at exponentially distributed waiting times. Due to the stochastic nature of these processes, the EP may significantly  vary  across different runs even for the same network realization.

To estimate the EP, we ran a total of \( n_t \) simulations, with multiple iterations performed across different network realizations. In each simulation, the infection rate was set to \( \beta_0(1 - \xi) \) during the interval \( t_0 \leq t \leq t_0 + T \), and to \( \beta_0 \) otherwise. The onset time \( t_0 \) was chosen to be several times the system’s relaxation time $t_r$ to make sure the system entered the long-lived metastable state prior to the initiation of the lockdown. The duration \( T \) was selected such that \( t_0 + T \) remained well below the MTE. Each simulation was propagated until either extinction occurred or a final cutoff time \( t_f\) was reached, such that \(t_f-t_0 \gg T \). We recorded the number of simulations that resulted in extinction, \( n_e \), and those that did not, \( n_s \), with \( n_t = n_e + n_s \). \textcolor{black}{The probability of extinction in the immediate aftermath of the lockdown, \( \mathcal{P}_a\), which we henceforth denote as \( \mathcal{P}\), is then computed as \( \mathcal{P} = n_e / n_t \).  While $\mathcal{P}$ depends on the lockdown parameters $\xi$ and $T$, it is insensitive to the specific choices of $t_0$ (as long as $t_r\!\ll\! t_0\!\ll\! \tau$) and $t_f$ (as long as $t_f\ll \tau$), see Appendix B. Finally, we ran a sufficient number of realizations per each parameter set such that the error in $\mathcal{P}$ was $< 10\%$.}

\section{\label{sec:results}  Results}
At this point we apply our numerical method to explore networks with homogeneous, bimodal, and gamma degree distributions across a range of parameters: \( T \), \( \xi \), \( \epsilon \), and \( \alpha \). These different network types allow us to examine distinct topological scenarios: homogeneous networks represent cases where topology plays no role; bimodal networks capture symmetric degree distributions; and gamma networks capture more realistic degree heterogeneity observed in real-world systems \cite{NOVOZHILOV2008defdistribution,Neipel2020epidemic_waves}. 
Notably, although the bimodal network is a simplified, toy model,  it provides qualitatively similar results to other, more complex, networks with a symmetric degree distribution~\cite{hindes_2019_undirected_degree_hetro}. Therefore, we will henceforth use the bimodal network as a prototypical symmetric network, while the gamma network will be used as a prototypical model for asymmetric, skewed distributions~\cite{korngut_2025_we}.

\begin{figure}[h]
    \centering
    \includegraphics[width=.97\linewidth]{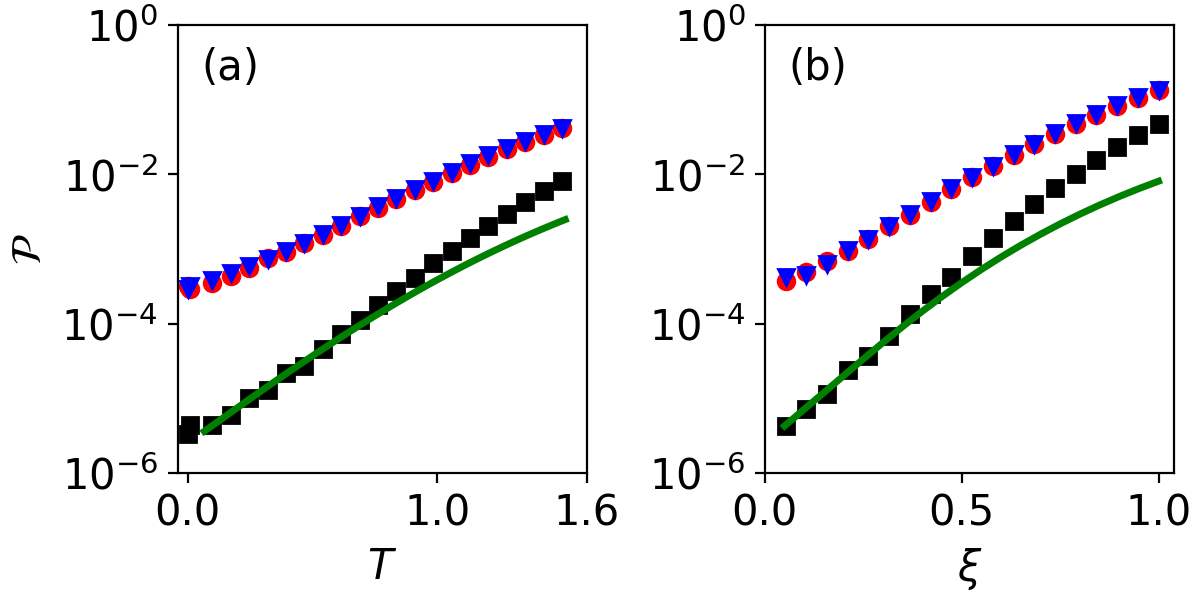}%
\vspace{-5mm}
    \caption{EP for random homogeneous (squares), bimodal (circles), and gamma (triangles) networks with \( N = 1000 \) and \( \langle k \rangle = 50 \). Here, $t_0=50$, $t_f=100$ and \( R_0 = 1.2 \). For the bimodal and gamma distributions we took a COV of \( \epsilon = 0.3 \). Panel (a) shows results for varying \( T \) with \( \xi = 1.0 \), while (b) corresponds to \( T = 2.0 \) with varying \( \xi \). The solid lines show the analytical results [Eq.~(\ref{eq:action_pe})] for the homogeneous case.} 
    \label{fig1}
\end{figure}

As a first step, we examined how the EP depends on the lockdown parameters \( T \) and \( \xi \) in homogeneous versus heterogeneous settings. Figure~\ref{fig1} shows the results of \( \mathcal{P} \) for random networks with homogeneous, bimodal, and gamma degree distributions. We also include a comparison with the analytical results of the homogeneous case,  which hold as long as $|\ln\mathcal{P}|\gg 1$, see Appendix A. Panels (a) and (b) demonstrate that increasing the duration or strength of the lockdown leads to a higher EP. Yet, in homogeneous networks, \( \mathcal{P} \) remains significantly lower than in heterogeneous ones. This result is encouraging, as it suggests that realistic heterogeneous networks are more susceptible to lockdown measures; that is, disease can be much more easily eradicated in heterogeneous settings.

Apart from the difference between homogeneous and heterogeneous settings, Fig.~\ref{fig1} also shows that different networks (bimodal and gamma) yield an almost identical EP, as long as $\epsilon$ is identical. This implies that  network heterogeneity, rather than the specific form of  degree distribution, is the key factor in determining the EP.  

\begin{figure}[h]
    \centering
    \includegraphics[width=.97\linewidth]{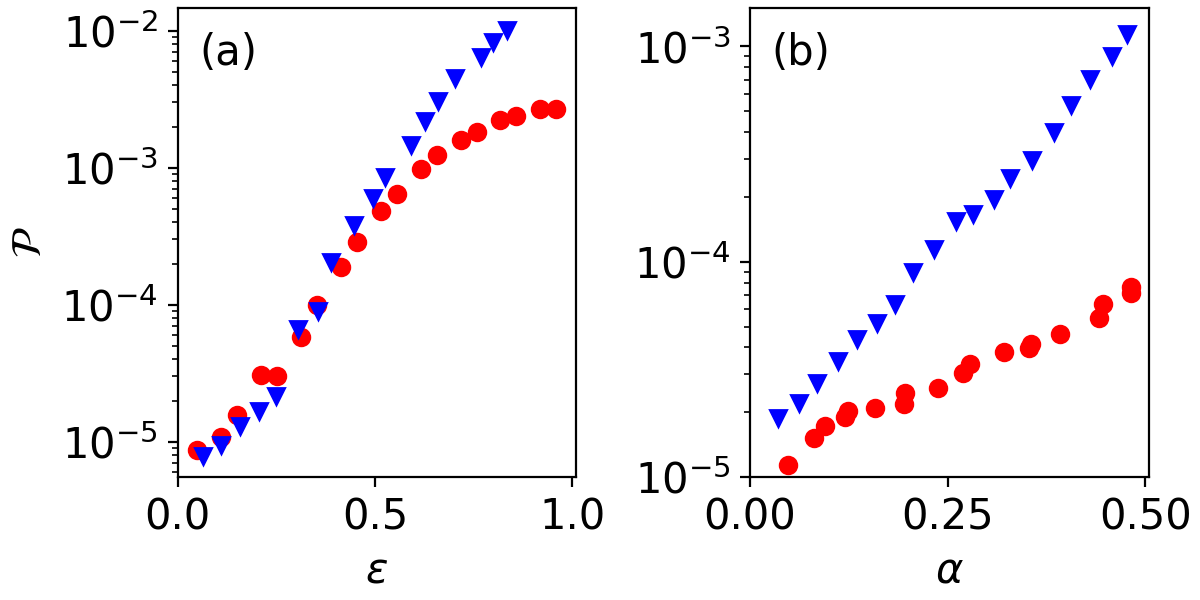}%
\vspace{-5mm}
    \caption{EP for bimodal (circles) and gamma (triangles) networks with \( N = 1000 \) and \( \langle k \rangle = 50 \). Here, $t_0=50$, $t_f=100$, \( T = 2.0 \), and \( \xi = 1.0 \). Panel~(a) shows random ($\alpha\!=\!0$) networks  with \( R_0 \!=\! 1.47 \) and varying \( \epsilon \), while panel (b) shows assortative networks with \( R_0 = 1.6 \), \( \epsilon = 0.5 \) and varying \( \alpha \).} 
    \label{fig2}
\end{figure}

\begin{figure}[h]
    \centering
    \includegraphics[width=0.86\linewidth]{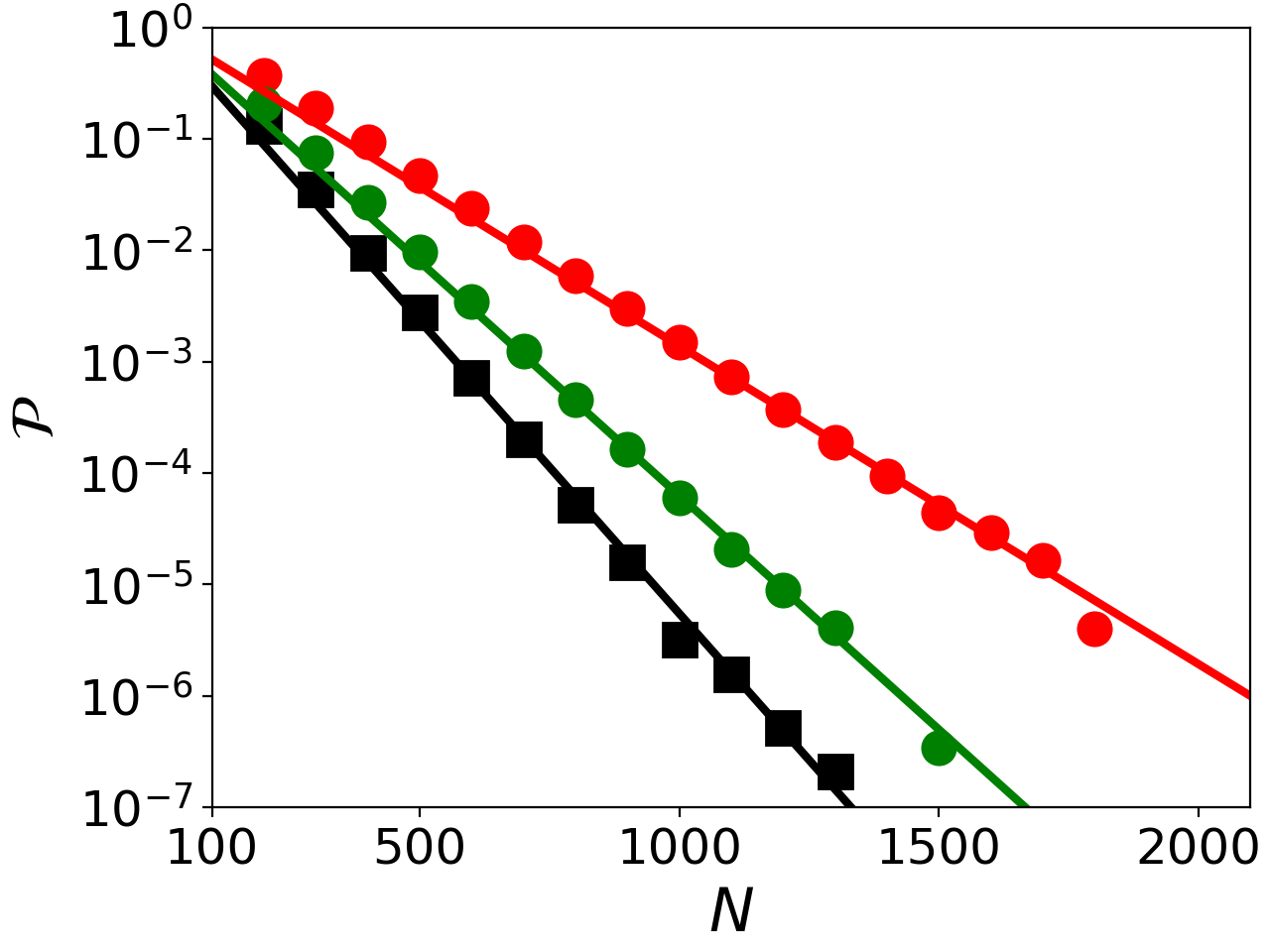}%
\vspace{-5mm}
   \caption{\textcolor{black}{EP versus $N$ for networks with $\left<k\right>=50$. Here, $t_0=50$, $t_f=100$, $T=2.0$, $\xi=1.0$ and $R_0=1.47$. Shown are homogeneous (squares) and bimodal (circles) networks, with red and green circles representing networks with $\epsilon=0.7$ and $\alpha=0.0$, and $\epsilon=0.25$ and $\alpha=0.25$, respectively. Solid lines represent a theoretical fit according to $e^{-N\Delta S}$. In the homogeneous case $\Delta S\simeq 0.0122$, within $6\%$ of the theoretical prediction, $0.013$ (see Appendix A), while for the bimodal networks with $\epsilon=0.7$ and $\epsilon=0.25$, a numerical fit yields $\Delta S\simeq 6.58\times10^{-3}$ and $\Delta S \simeq 9.67\times10^{-3}$, respectively.}} 
    \label{fig3}
\end{figure}

This attribute is further illustrated in panel (a) of Fig~\ref{fig2}, where \( \mathcal{P} \) is plotted as a function of \( \epsilon \) for both bimodal and gamma random networks. In both cases, \( \mathcal{P} \) increases with heterogeneity, and as long as $\epsilon<0.5$ the curves remain close. However, as $\epsilon$ is further increased, the EP plateaus for bimodal networks, whereas for gamma networks, $\mathcal{P}$ continues to rise sharply.
The discrepancy at large $\epsilon$ arises because in symmetric networks, $\epsilon$ is bound by $1$ as the standard deviation can be at most the size of the mean. For bimodal networks, as \( \epsilon \) approaches 1, the network becomes effectively split: half the nodes have degree $2k_0-1$, while the other half have a degree of $1$. \textcolor{black}{In Fig.~\ref{fig3} we demonstrate that the qualitative dependence of the EP on the system size $N$ for heterogeneous networks (both random or assortative)}  is similar to the homogeneous case: $\mathcal{P}\sim e^{-N \Delta S(\epsilon,\alpha,T,\xi)}$. Here, $\Delta S(\dots)$ is the action barrier, which depends on the network heterogeneity, assortativity, lockdown duration and strength (see Appendix A for details). Therefore, as $\epsilon$ approaches $1$ and the bimodal network becomes practically split, the effective system size becomes $N/2$, and the EP should approach the square root of the value at $\epsilon=0$, exactly as seen in Fig.~\ref{fig2}(a).  In contrast, in asymmetric networks, such as the gamma distribution, $\epsilon$ is not bounded. Here, for \( \epsilon \gtrsim 1 \), a few high-degree hubs dominate transmission. These hubs, while efficient in spreading the disease, also make the network more susceptible to extinction via large  fluctuations~\cite{korngut_2025_we,korngut2025_assortative}. Therefore, as $\epsilon$ grows, the EP continues to grow without saturating.

Another key quantity  which strongly influences the EP is the network assortativity.  In Fig~\ref{fig2}(b), we plot \( \mathcal{P} \) as a function of \( \alpha \) for bimodal and gamma  networks while keeping \( \epsilon \) fixed. The results indicate that as assortativity or degree-degree correlations increase, the EP grows in both network types. However, the effect is more pronounced in gamma-distributed networks. The reason is that as $\alpha$ is increased, a high-degree clique is formed in the networks. However, in gamma networks, there are relatively more hubs than in bimodal networks with the same COV. Therefore, once the high-degree clique becomes susceptible again, the effective reproductive number decreases in a sharper manner in a gamma network compared to a bimodal network, and thus, the EP is higher for gamma networks, for the same $\alpha$ values.

\begin{figure}[h]
      \centering
  \hspace{-.5cm}\includegraphics[width=1.05\linewidth]{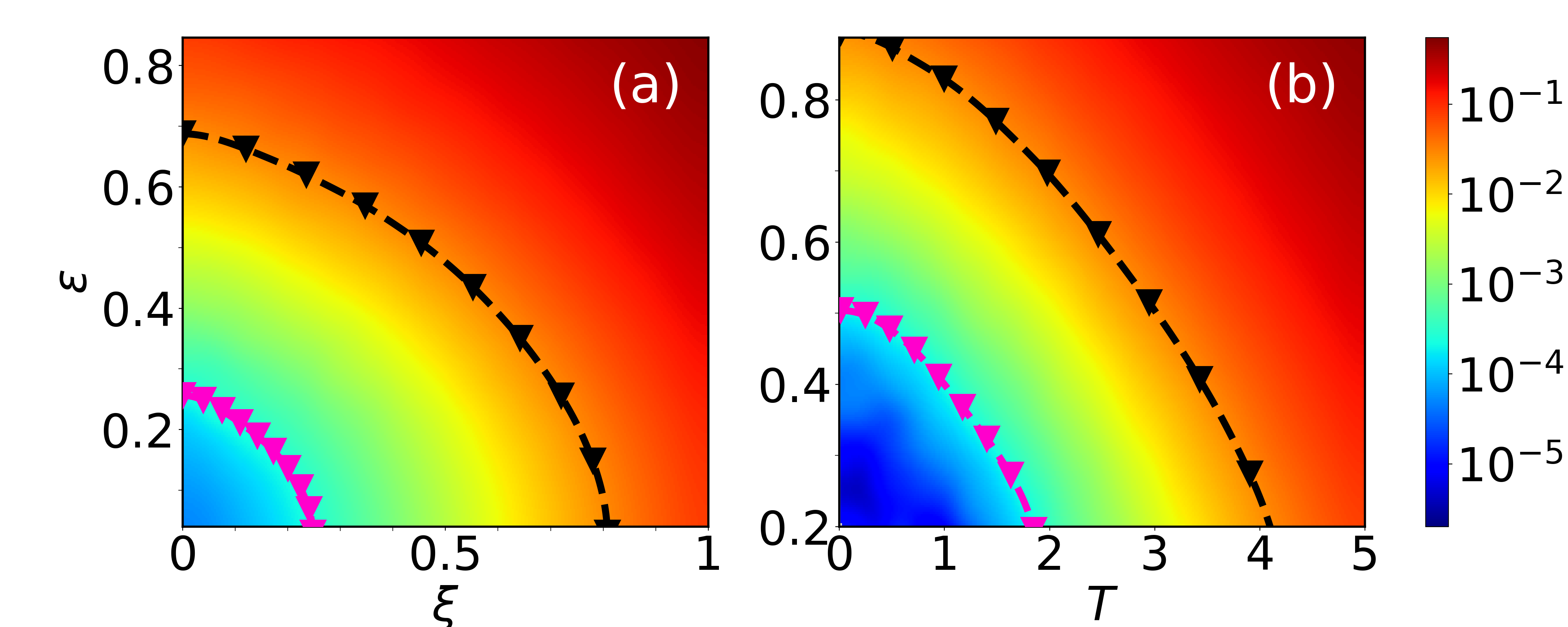}%
\vspace{-4mm}
    \caption{Shown are heatmaps of the EP,  on a logarithmic color scale (shared colorbar), along
  with contour lines of equal EP, $\mathcal{P}_T = \mathcal{F}\mathcal{P}_0$, computed from KMC simulations on random ($\alpha=0$) gamma-distributed networks. Here, $N = 1000$, $R_0 = 1.24$, $t_0=50$, $t_f=100$ and $\langle k \rangle = 50$. In (a), $\mathcal{P}$ is plotted as a function of $\xi$ and $\epsilon$ for  fixed duration $T = 2.0$, while in (b) $\mathcal{P}$ is plotted as a function of $T$ and $\epsilon$ for fixed $\xi = 1.0$. Magenta and black triangles and dashed lines correspond to isoprobability levels $\mathcal{F} = 10^4$ and $\mathcal{F} = 10^6$, respectively. The reference value $\mathcal{P}_0
= 2.96 \times 10^{-8}$ corresponds to the EP without lockdown in a homogeneous network, obtained from $10^{10}$ realizations~\cite{realizations}, whereas each $\mathcal{P}$ value was calculated using $10^8$ realizations.}
    \label{fig4}
\end{figure}

Overall, Fig.~\ref{fig2} demonstrates qualitative similarities between the effect of increasing $\epsilon$ and $\alpha$ on the EP. Namely, the lockdown becomes more effective as either $\epsilon$ or $\alpha$ increase\textcolor{black}{, regardless of the explicit choice of $N$ and $R_0$}. In addition, in both cases, the contrast between symmetric and asymmetric networks becomes evident and significant at high values of these parameters.

The results in Figs.~\ref{fig1}-\ref{fig3} raise a key question: can combinations of \( \epsilon \), \( \alpha \), \( T \) and \( \xi \) be found such that the EP remains unchanged? \textcolor{black}{More precisely, since network features are not freely tunable in practice,  what combination of changes in the dynamical parameters---lockdown strength $\xi$ and duration $T$---is required to achieve a given EP for a network with fixed structural properties, characterized by degree heterogeneity and assortativity?}


To address this, we conducted extensive numerical simulations on gamma-distributed random networks with linearly spaced values of \( \epsilon \) in the range \( 0.01 \leq \epsilon \leq 1.0 \) and varying lockdown strengths \( 0 \leq \xi \leq 1 \). \textcolor{black}{This produced heatmaps of $\mathcal{P}$ (with logarithmic colorbar), see Fig.~\ref{fig4}, for networks with fixed \( T \) and \( R_0 \) (Fig.~\ref{fig4}a), and fixed $\xi$ and $R_0$ (Fig.~\ref{fig4}b).
The overlaid isoprobability lines, denoted by \( \mathcal{P}_T \),} were defined relative to a homogeneous network  \( \mathcal{P}_0 \) in the absence of a lockdown (with \( \xi = \epsilon = 0 \)), via \( \mathcal{P}_T = \mathcal{F}\mathcal{P}_0 \), with \( \mathcal{F} \) being a prescribed numerical factor. In Fig.~\ref{fig4}(a) we show combinations of \( \epsilon \) and \( \xi \) that yield the target EP, \( \mathcal{P}_T \), which equals $10^4$ and $10^6$ times the unperturbed value in a homogeneous setting. While the choice of the factor $\mathcal{F}$ was arbitrary, we see that  as heterogeneity increases, the lockdown strength must decrease to produce the same effect. 
Similarly, in Fig.~\ref{fig4}(b) we show isoprobability curves for gamma-distributed networks with \( 0.01 \leq \epsilon \leq 1.0 \) and varying lockdown durations \( 0 \leq T \leq 4 \), for fixed \( \xi \) and \( R_0 \), and for the same $\mathcal{F}$ values as in  (a). Here, one can see that as heterogeneity increases, shorter lockdown durations must be taken to obtain the same effect, i.e., to maintain the same EP.

Figure~\ref{fig4} suggests that as network heterogeneity increases, the disease is more susceptible to extinction under lockdown interventions, reducing the level of effort required to achieve a given EP. These results are consistent with previous studies showing that increasing heterogeneity in undirected networks lowers the MTE, thereby making disease eradication more likely \cite{hindes_2019_undirected_degree_hetro,clancy_2018_persistence,korngut_2022_direction,korngut_2025_we,korngut2025_assortative}.

Since real-world networks are rarely random, and often exhibit either assortative or disassortative mixing, we wanted to check whether the results in Fig.~\ref{fig4} can be generalized to \textcolor{black}{more realistic scenarios} with positive or negative degree-degree correlations. As shown in Fig.~\ref{fig2},  even networks with the same level of heterogeneity can display different EPs due to their assortative structure. To study this effect, we adopted the same approach as in the heterogeneous case, and generated multiple gamma-distributed networks with varying assortativity, mapping the state space, and identifying all combinations of lockdown parameters and \( \alpha \) values that yield the same EP.  

\begin{figure}[h]
    \centering
    \hspace{-0.2cm}\includegraphics[width=1.02\linewidth]{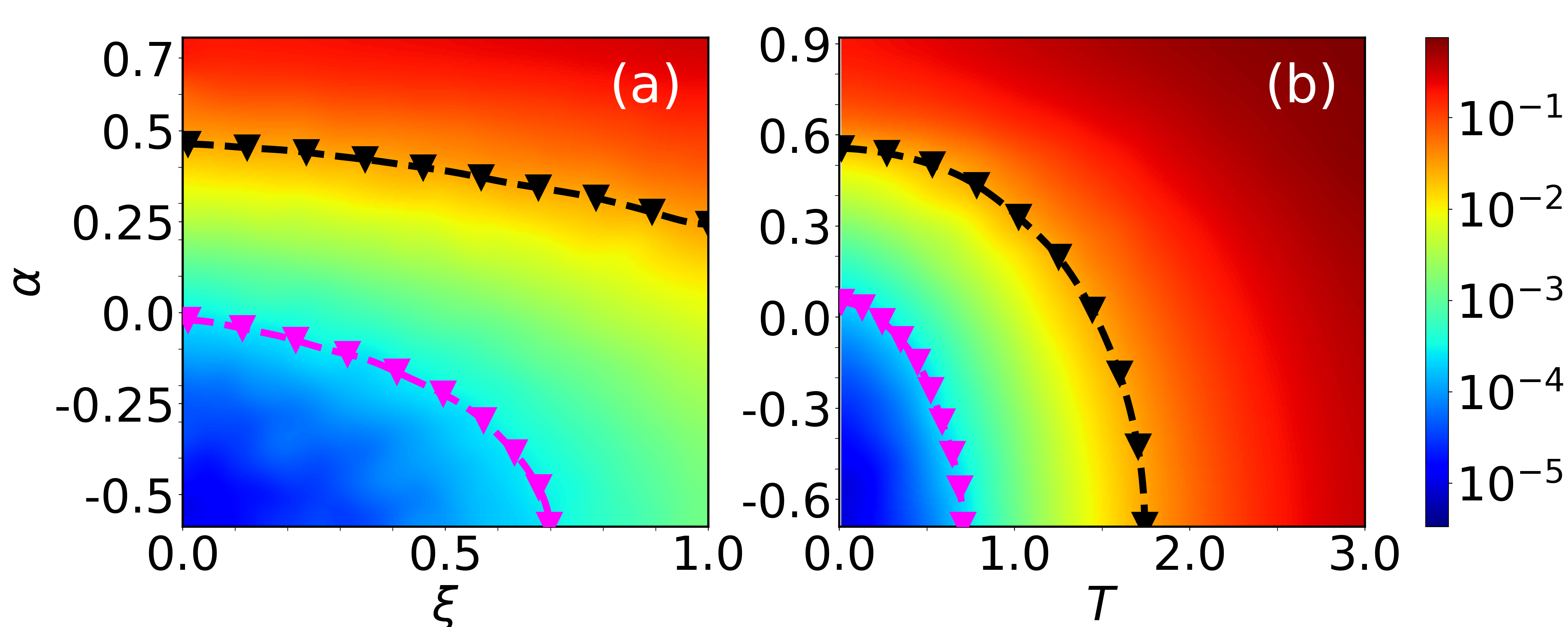}%
\vspace{-4mm}
    \caption{
Shown are heatmaps of the EP, on a logarithmic color scale (shared colorbar), along with contour lines of equal EP, $\mathcal{P}_T\!=\!\mathcal{F}\mathcal{P}_0$, computed from KMC simulations on (dis)assortative gamma-distributed networks. Here, $N = 1000$, $R_0 \!=\! 1.24$, $t_0=50$, $t_f=100$, $\epsilon\!=\!0.5$ and $\langle k \rangle \!=\! 50$. In (a) $\mathcal{P}$ is plotted as a function of $\xi$ and $\alpha$ for a fixed duration $T = 1.0$, while in (b), $\mathcal{P}$ is shown versus $T$ and $\alpha$, for a fixed $\xi=1.0$. Magenta and black triangles and dashed lines correspond to isoprobability levels $\mathcal{F}=10^4$ and $\mathcal{F}=10^6$, respectively. The reference value $\mathcal{P}_0=2.96\times10^{-8}$ corresponds to the EP without lockdown in a homogeneous network, obtained by performing $10^{10}$ realizations~\cite{realizations}, whereas each $\mathcal{P}$ value was calculated using $10^8$ realizations.}
    \label{fig5}
\end{figure}

\textcolor{black}{In Fig.~\ref{fig5} we show heatmaps of $\mathcal{P}$ (with logarithmic colorbar) for assortative and disassortative gamma-distributed networks as function of $\xi$ and $\alpha$ (Fig.~\ref{fig5}a), and $T$ and $\alpha$ (Fig.~\ref{fig5}b). We also overlay isoprobability contour lines of \( \mathcal{P}_T = \mathcal{F}\mathcal{P}_0 \), with the same $\mathcal{F}$ values as in Fig.~\ref{fig4}}. These results indicate that as assortativity increases, the values of \( T \) and \( \xi \) required to achieve a fixed EP decrease,  demonstrating that, much like heterogeneity, assortativity enhances the likelihood of disease eradication under intervention. Notably, this qualitative similarity between the role of heterogeneity and assortativity is consistent with previous studies in static networks~\cite{korngut2025_assortative}.  

The results in both Figs.~\ref{fig4} and \ref{fig5} are encouraging, as they suggest that in structured, realistic environments, the lockdown measures needed to eliminate an outbreak may be substantially less demanding than predicted by simpler models, making such interventions much more feasible and useful in practice.

\section{Discussion \label{sec:discussion}}
We have studied the impact of  abrupt environmental changes on the long-lived endemic state within the SIS model of epidemics. Such a temporary environmental switch may dramatically expedite disease clearance, even though the perturbed state's duration is short. While previous works have addressed such effects in periodically varying environments or in deterministic settings ~\cite{Keeling_2008_stochastic_dynamics,shaw2008_adaptive_networks,assaf2009population,bacaer2015stochastic,Israeli2020switching,Atias2025_quarantine}, the effect of a temporary lockdown has not been explored in the context of a lingering disease in realistic heterogeneous topologies.

Here, we have systematically studied the impact of a temporary lockdown on the extinction  probability (EP) of a long-lived disease, on degree-heterogeneous (dis)assortative networks. The lockdown was modeled as a sharp decrease in the infection rate, with a given magnitude and for a prescribed duration. The resulting EP was the fraction of realizations that got extinct in the immediate aftermath of the lockdown. 

First, we showed that the EP grows as the duration or magnitude of the lockdown are increased.  
Next, we showed that for a given lockdown, the EP can be greatly enhanced by either increasing the network's heterogeneity or its assortativity. Enhancing the network's heterogeneity gives rise to an increasing number of hubs, which upon recovery, lower the reproductive number significantly and thereby expedite disease clearance and increase the EP. A similar effect occurs upon increasing the network's assortativity, giving rise to high-degree network cliques, which also greatly expedite disease clearance. 
The key consequence of these results is that, as degree heterogeneity or assortativity increases, milder lockdowns---with shorter durations or weaker magnitude---suffice to achieve a desired EP.

\textcolor{black}{We then computed heatmaps of the EP as a function of the lockdown and network parameters and overlaid equal-EP contour lines}. This allowed us to study the interplay between lockdown magnitude and duration, and extent of network heterogeneity and assortativity. Notably, we showed that in structured, realistic environments,
the lockdown measures required to eradicate a disease
may be far less demanding than predicted by
simpler models. This means that such interventions are much more
effective, which may revolutionize health policies  in realistic scenarios.

While we focused on lockdowns that affect the general population, targeted interventions---where only specific individuals, such as high-degree nodes, experience a sharp drop in their infection rate---present an interesting direction for future research~\cite{Hacohen2022}. Another promising avenue is the study of repeated lockdowns in such networks, which may influence disease dynamics differently from a single prolonged one~\cite{Meidan2021}. 
Lastly, although we modeled lockdowns as abrupt reductions in the infection rate, other modeling approaches may be worth exploring; e.g., scenarios where the infection rate changes gradually or follows a non-symmetric profile, or when the network topology itself undergoes sudden structural changes.

\section{Acknowledgments}
\noindent EK and MA acknowledge support from  ISF
grant~531/20.

\section*{Appendix A: well-mixed case}
\setcounter{figure}{0}
\setcounter{equation}{0}
\renewcommand{\thefigure}{A\arabic{figure}}
\renewcommand{\theequation}{A\arabic{equation}} 

Here we consider a well-mixed (or homogeneous) population, and derive the extinction probability (EP) in the aftermath of a temporary lockdown, using a semiclassical approximation valid as long as $|\!\ln\mathcal{P}|\!\gg\! 1$ \textcolor{black}{(see below)}.

In a well-mixed population, each individual interacts with all others. Thus, one can write the infection and recovery  reactions in the following way:
$I\xrightarrow{W_{+}} I+1$ and $I\xrightarrow{W_{-}} I-1$,
where  $W_{+}\!=\!\beta I(N\!-\!I)/N$ is the infection rate, and $W_{-}\!=\!\gamma I$ is the  recovery rate. In the limit of an infinite population, one can ignore demographic noise and write the  mean-field rate equation 
\begin{equation}
\label{eq:rate}
    \dot{x}=\beta (t)x\left(1-x\right)-\gamma x,
\end{equation}
where $x=I/N$ is the  fraction  of infected. This equation has two steady-state solutions: an endemic stable state at $x^{*}=1-1/R_0$, where $R_0=\beta/\gamma$ is the basic reproductive number, and an extinct unstable state at $x=0$~\cite{Pastor-Satorras_2015_anneald_net}.

In reality, however, the population is finite and thus, demographic noise must be incorporated. Since the extinct state is absorbing, this noise renders the endemic state metastable~\cite{dykman_1994_Large_fluctuations,assaf_2010_Extinction,assaf_2017_WKB}. As a result, we consider the probability $P(I,t)$ of having $I$ infected at time $t$, whose time evolution satisfies the following master equation~\cite{gardiner2009stochastic}
\begin{eqnarray}
    &&\hspace{-3mm}\frac{\partial P(I,t)}{\partial t} = W_+(I-1,t)P(I-1,t)-W_+(I,t)P(I,t)\nonumber\\
    &&\hspace{3mm}+W_{-}(I+1)P(I+1,t)-W_{-}(I) P(I,t).
\label{eq:master_generic}
\end{eqnarray}
Our goal is to find the EP, $\mathcal{P}\!=\!P(I\!=\!0,t)$, in the immediate aftermath of the lockdown. In the limit of large population size, $N\gg 1$, after a short transient the system  enters a long-lived metastable endemic state. The endemic state can be described as a slowly decaying distribution,  \( P(I, t) = \pi(I)\,e^{-t/\tau} \), where $I=1,\dots,N$. Here, \( \tau \) is the (exponentially large) MTE, and \( \pi(I) \) is the normalized QSD---the shape of the metastable state~\cite{dykman_1994_Large_fluctuations,assaf_2010_Extinction,assaf_2017_WKB}. Finding the EP in the time-independent case is simple, since $P(0,t)=1-\sum P(n>0,t)=1-e^{-t/\tau}$. As a result, at times $t\ll \tau$, the EP grows linearly in time at an exponentially slow rate, and satisfies $\mathcal{P}(t)\simeq t/\tau\sim \tau^{-1}$.

To estimate the EP under a temporary lockdown, we first show how the MTE can be found in the time-independent case. To do so, we employ the WKB (Wentzel–Kramers–Brillouin) method to approximate the QSD, and write $\pi(I) \equiv \pi(x) \sim e^{-N \mathcal{S}(x)}$ where \( \mathcal{S}(x) \) is the action function~\cite{dykman_1994_Large_fluctuations,assaf_2010_Extinction,assaf_2017_WKB,hindes_2016_paths}. Plugging this into  Eq.~(\ref{eq:master_generic}) and neglecting the exponentially-small term on the left hand side, we arrive at a stationary Hamilton-Jacobi equation \( H_0(x, \partial_{x} \mathcal{S}) = 0 \) with Hamiltonian
\begin{equation}
    \label{eq:hamiltionian}
    H_0(x,p)\!=(1-e^{-p})x\left[e^{p}R_{0}(1-x)-1\right].
\end{equation}
Here, $p\!=\!\partial_{x}\mathcal{S}$ is the associated momentum, and we have added a subscript of $0$ indicating an unperturbed Hamiltonian. To find the action,  one has to find the optimal path to extinction---a heteroclinic trajectory connecting the metastable endemic state and extinct state. Demanding that \( H_0[x, p_0(x)] = 0 \) and using Eq.~\eqref{eq:hamiltionian} we find
\begin{equation}
\label{eq:path0}
    p_0(x) = -\ln\left[R_0\left(1-x\right)\right].
\end{equation}
Alternatively, the optimal path can be found by solving the Hamilton's equations, \( \dot{x} = \partial_{p} H \) and \( \dot{p} = -\partial_{x} H \). 
Given the optimal path, the  MTE is given  by~\cite{ovaskainen2010stochastic,assaf_2017_WKB}
\begin{equation}
\label{eq:action}
     \hspace{-2mm}\tau\sim e^{N\Delta\mathcal{S}}\!,\;\;\Delta\mathcal{S}\!=\!\int_{0}^{\infty}\!\![p\dot{x}-H(x,p)]dt=\int_{x^*}^0\!\! p_0(x)dx.
\end{equation}
Here, we have denoted the action barrier for extinction by \( \Delta \mathcal{S} \), and used the fact that in the time-independent case, $H_0(x,p)=0$. Substituting \( p_0(x) \) from Eq.~(\ref{eq:path0}) into 
Eq.~\eqref{eq:action} yields \(\Delta S_0 = \ln(R_0) + 1/R_0 - 1 \), leading to $\tau \sim e^{N \Delta \mathcal{S}_0}= e^{N (\ln R_0+1/R_0-1)}$,
up to pre-exponential corrections~\cite{assaf_2010_Extinction}. Thus, the EP in the absence of a lockdown is indeed exponentially small and satisfies $\mathcal{P}_b\sim e^{-N\Delta\mathcal{S}_0}$.

Having computed the EP in the time-independent case, we now turn to study the lockdown scenario. Here, the infection rate becomes explicitly time-dependent and is given by Eq.~(\ref{eq:beta_quant}). 
We will assume that the lockdown is sufficiently strong (or long) such that the EP in the aftermath of the lockdown $\mathcal{P}_a$ satisfies $\mathcal{P}_a\gg \mathcal{P}_b$. The increase in EP due to the lockdown is therefore given by \( \Delta \mathcal{P} \equiv \mathcal{P}_a - \mathcal{P}_b \simeq \mathcal{P}_a \), and under the WKB approximation takes the form,
$\Delta \mathcal{P} \sim e^{-N \Delta \mathcal{S}}$~\cite{assaf2009population,Israeli2020switching}. We will now compute this action barrier in the presence of a lockdown.
\vspace{-4mm}
\begin{figure}[h]
    \centering
    \includegraphics[width=0.8\linewidth]{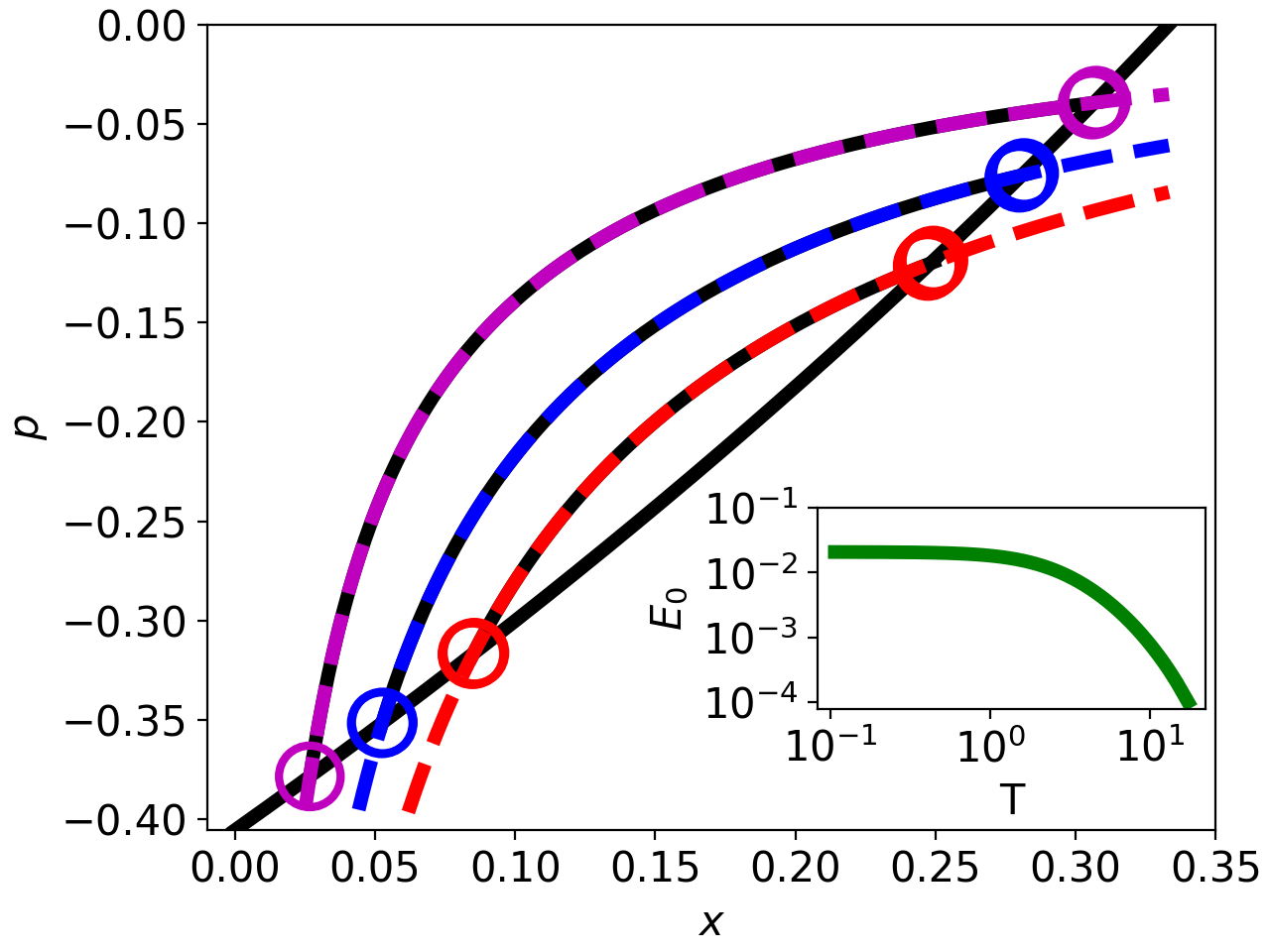}%
\vspace{-5mm}
    \caption{Extinction path \( p \) versus \( x \), for \( R_0 = 1.5 \) and \( \xi = 0.5 \). Solid lines show the unperturbed  path~\eqref{eq:path0}, while dashed lines show the paths during lockdown~\eqref{eq:pathenergy1d} for durations \( T = 1.5, 2.5, 4 \) (bottom to top) for which  $E_0\simeq 0.01584$, $0.01106$, and $0.00615$. Circles mark the intersections between the perturbed and unperturbed segments. Inset shows a log-log plot of $E_0$ versus $T$ for $R_0=1.5$ and $\xi=0.5$.}
    \label{fig6}
\end{figure}

The action barrier to extinction in the aftermath of the environmental
change can be found by integrating along the optimal path to
extinction, $\Delta \mathcal{S}=\int_0^{\infty}[p\dot{x}-H(x,p)]dt$. This  trajectory starts at
the endemic saddle point $(x, p) = (x^*, 0)$ well before the lockdown
has been applied, and ends at the extinction saddle point $(0, -\ln R_0)$
well after the lockdown has been lifted. Yet, despite
the explicit time dependence of the environment in this case,
the integration along the optimal path can be done analytically.
This is because the infection rate here changes from
one constant value to another, and thus there are now two different
time-independent Hamiltonians, both of which
are integrals of motion. These are the unperturbed Hamiltonian
at times $t < t_0$ and $t > t_0 + T$, $H(x, p) = H_0(x, p)$ [Eq.~(\ref{eq:hamiltionian})], and the perturbed Hamiltonian, $H(x, p) = H_p(x, p)$,
at times $t_0 < t < t_0 + T$, which reads~\cite{assaf_2010_Extinction}:
    $H_p(x,p)\!=(1-e^{-p})x\left[e^{p}R_{0}(1-\xi)(1-x)-1\right]$.
    Yet, while the optimal path before and after the lockdown, \( p_0(x) \), is determined by the zero-energy trajectory of the Hamiltonian, during the lockdown the energy takes some constant value \( E_0 \), which is generally nonzero. Solving \( H_p(x, p) = E_0 \) for the perturbed segment yields,
\begin{equation}
\label{eq:pathenergy1d}
p_p(x;E_0)=-\ln\!\left\{\left[\Psi\!+\!\sqrt{\Psi^2\!-\!4R_0x^2(1\!-\!x)(1\!-\!\xi)}\right]\!/(2x)\right\}\!,
\end{equation}
where $\Psi=E_0+x+R_0(1-x)x(1-\xi)$. Here, the energy $E_0$ can be implicitly determined by demanding that the duration of the lockdown trajectory be exactly $T$~\cite{assaf2009population,Israeli2020switching}
\begin{eqnarray}
    \label{eq:integral_T}
    T&=& \int_{0}^{ T}\!\!dt=\int_{x_{+}^{p}\left(E_0\right)}^{x_{-}^{p}\left(E_0\right)}\!\!\frac{dx}{\dot{x}\left[x,p_p\left(x;E_0\right)\right]}.
\end{eqnarray}
The integral boundaries $x_{-}^{p}\left(E_0\right)$ and $x_{+}^{p}\left(E_0\right)$ are the lower and upper intersection points between the unperturbed [Eq.~\eqref{eq:path0}] and  perturbed [Eq.~\eqref{eq:pathenergy1d}] optimal paths (see Fig.~\ref{fig6}), and are given by
\begin{equation}\label{xpm}
    x_{\pm}^{p}(E_0)=(x^*/2)\left[1\!\pm\!\sqrt{1\!-\!4E_0(1\!-\!x^*)/(\xi x^*{}^2)}\right],
\end{equation}
where $x^*=1-1/R_0$ is the unperturbed endemic state.

Plugging the Hamilton's equation for $\dot{x}$ during the perturbed trajectory, $dx/dt=\partial_p H_p(x,p)=e^p R_0(1-x)x(1-\xi)-xe^{-p}$ into Eq.~(\ref{eq:integral_T}), we find 
\begin{equation}\label{TE0}
    T=\int_{x_{+}^{p}\left(E_0\right)}^{x_{-}^{p}\left(E_0\right)}\!\!\left[\sqrt{\Psi^2-4R_0x^2(1-x)(1-\xi)}\right]^{-1}dx,
\end{equation}
where $\Psi$ is defined below Eq.~(\ref{eq:pathenergy1d}), and depends on $E_0$.

Figure~\ref{fig6} shows the extinction paths for different values of \( T \), where we  used Eq.~(\ref{TE0}) to compute $E_0$ for each $T$. 
As \( T \) increases, the deviation from the unperturbed trajectory~\eqref{eq:path0} increases, 
and the intersection points between  Eqs.~\eqref{eq:path0} and ~\eqref{eq:pathenergy1d}  
define a longer segment over which the trajectory follows Eq.~\eqref{eq:pathenergy1d}. 
This extended segment increases the lockdown's influence on the action barrier; as $T$ increases, $\Delta \mathcal{S}$ decreases and EP increases.

Putting everything together, we can compute the action barrier to extinction, $\Delta\mathcal{S}$, as follows:
\begin{eqnarray}
    \label{eq:action_pe}
    &&\hspace{-5mm}\Delta \mathcal{S}
        =\int _{0}^{\infty} \!\![p(x)\dot{x} -H]dt=\int_{x^*}^{0}\!p(x)dx+\int_{t_0}^{t_0+T}\!\!H_p dt\nonumber\\ 
    &&=\int_{x^*}^{0}\!\!\!\!p_0(x)dx\!-\!\int_{x_{+}^{p}(E_0)}^{x_{-}^{p}(E_0)}\!\!\!\![p_0(x)-p_p(x;E_0)]dx\!-\!\int_{t_0}^{t_0+T}\!\!\!\!\!\!\!\!\!\!\!\!H_p dt\nonumber\\
    &&=\Delta \mathcal{S}_0 -E_0 T -\int_{x_{+}^{p}(E_0)}^{x_{-}^{p}(E_0)}\!\!\left[ p_0(x)-p_p\left(x;E_0\right) \right] \textcolor{black}{dx},    
\end{eqnarray}
where $p_0(x)$, $p_p(x;E_0)$ and $x_{\pm}^{p}(E_0)$ are given by Eqs. (\ref{eq:path0}), (\ref{eq:pathenergy1d}), and (\ref{xpm}) respectively, and $E_0$ can be found from Eq.~\eqref{TE0}. In Fig.~\ref{fig1} we include a comparison between the numerical results in the homogeneous case and our analytical result, $\mathcal{P}\sim e^{-N\Delta\mathcal{S}}$ with Eq.~(\ref{eq:action_pe}). Notably, the agreement is excellent as long as $N\Delta\mathcal{S}\gg 1$.

\vspace{-0.2cm}
\section*{Appendix B: dependence of 
extinction probability on cutoff time}
\setcounter{figure}{0}
\setcounter{equation}{0}
\renewcommand{\thefigure}{B\arabic{figure}}

\begin{figure}[h]
    \centering
    \includegraphics[width=0.76\linewidth]{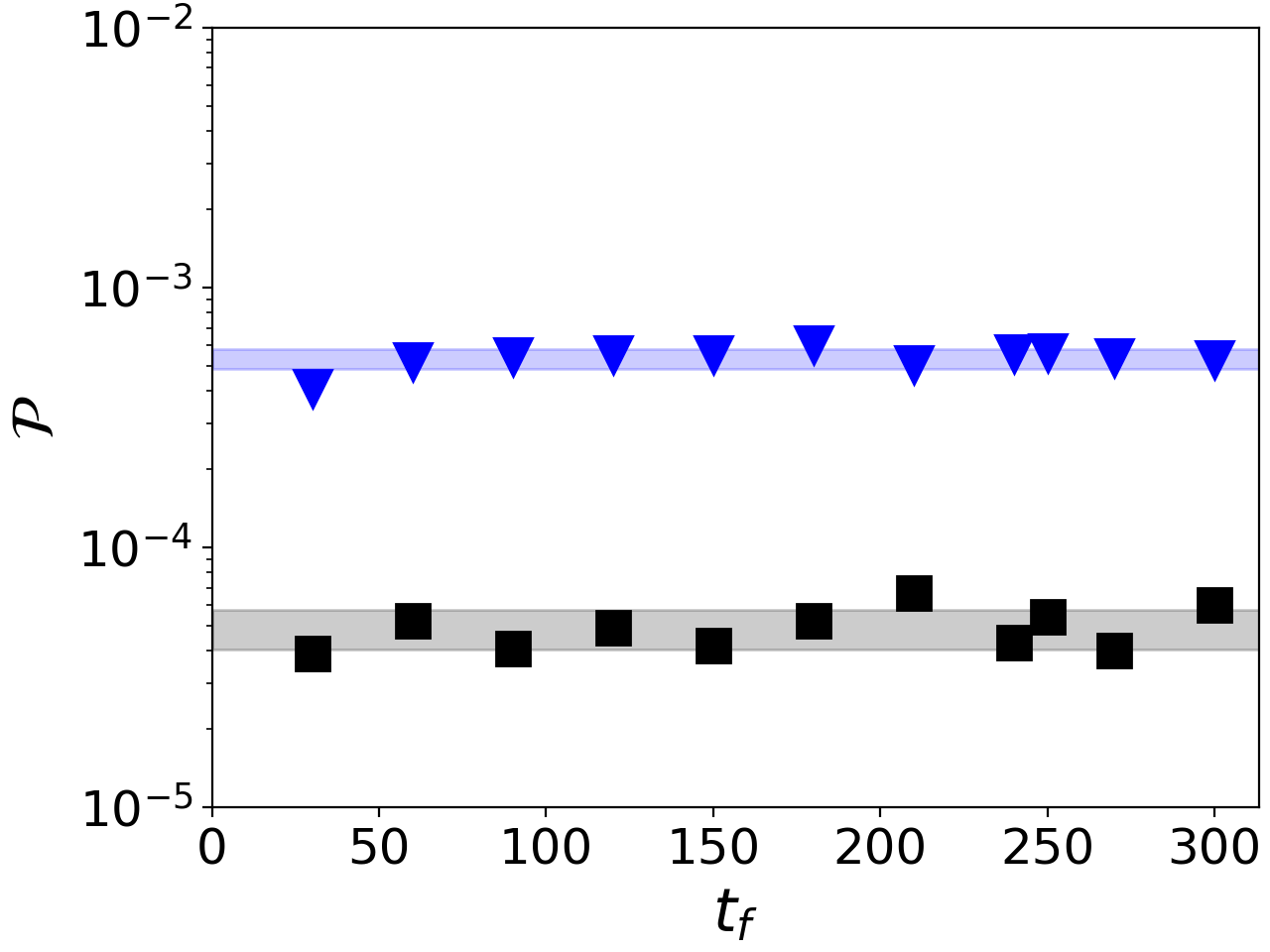}%
\vspace{-5mm}
    \caption{\textcolor{black}{EP versus the final cutoff time \( t_f \) for random homogeneous (squares) and gamma-distributed (triangles) networks with $N=1000$, $\left<k\right>=50$, and $\epsilon=0.3$ for the gamma network. Here, $t_0=20.0$ ,$T=2.0$, $\xi=1.0$ and $R_0=1.4$. The shaded regions around the respective means show one standard deviation above and below. The MTE for these networks is at least an order of magnitude higher than $t_f$.}}
    \vspace{-0.3cm}
    \label{fig7}
\end{figure}

\textcolor{black}{Here, we examine the dependence of the EP on the final cutoff time $t_f$. In Fig.~\ref{fig7} we plot the EP as a function of  $t_f$ for two random networks: a homogeneous network and a gamma-distributed network, where  $t_f$ is varied by an order of magnitude from $30$ to $300$ (with $t_0=20$ and $T=2.0$). The figure clearly demonstrates that the EP does not depend on the cutoff time, provided that the cutoff time $t_f$ is much smaller than the respective MTE.}

\bibliography{hetbib}

\end{document}